\documentclass[aps,prb,twocolumn,floatfix,showkeys]{revtex4}
\usepackage{times}
\usepackage{graphicx}
\usepackage{epsfig}
\usepackage{amsmath}
\usepackage{amssymb}
\usepackage[version=3]{mhchem}
\usepackage{nicefrac,xfrac}

\begin{document}
	\title{Quantitative three-dimensional local order analysis of nanomaterials through electron diffraction}
	
	\author{Ella Mara Schmidt$^\ast$}
		\affiliation{Faculty of Geoscience and MAPEX Center for Materials and Processes, University of Bremen, Bremen, Germany.}
\affiliation{MARUM-Center for Marine Environmental Sciences, University of Bremen, Bremen, Germany.}
\affiliation{Department of Chemistry, University of Oxford, Inorganic Chemistry Laboratory, South Parks Road, Oxford OX1 3QR, U.K.}
\affiliation{$\ast$ Corresponding author e-mail: ella.schmidt@uni-bremen.de}
		
	\author{Paul Benjamin Klar}
		\affiliation{Faculty of Geoscience and MAPEX Center for Materials and Processes, University of Bremen, Bremen, Germany.}
	\affiliation{Institute of Physics of the Czech Academy of Sciences, Prague, Czechia.}
	
		\author{Ya\c sar Krysiak}
\affiliation{Institute of Inorganic Chemistry, Leibniz University Hannover, Hannover Germany.}
		\affiliation{Institute of Physics of the Czech Academy of Sciences, Prague, Czechia.}
		\author{Petr Svora}
			\affiliation{Institute of Physics of the Czech Academy of Sciences, Prague, Czechia.}

		\author{Andrew L. Goodwin}
\affiliation{Department of Chemistry, University of Oxford, Inorganic Chemistry Laboratory, South Parks Road, Oxford OX1 3QR, U.K.}
	
	\author{Lukas Palatinus}
		\affiliation{Institute of Physics of the Czech Academy of Sciences, Prague, Czechia.}

	\date{\today}
	
	\newcommand{\abstractText}{\noindent

	}

%	\pacs{test test}

			\begin{abstract}
			Structure-property relationships in ordered materials have long been a core principle in materials design. However, the intentional introduction of disorder into materials provides structural flexibility and thus access to material properties that are not attainable in conventional, ordered materials. To understand disorder-property relationships, the disorder - i.e., the local ordering principles – must be quantified. Correlated disorder can be probed experimentally by diffuse scattering. The analysis is notoriously difficult, especially if only powder samples are available. Here, we combine the advantages of three-dimensional electron diffraction - a method that allows single crystal diffraction measurements on sub-micron sized crystals – and three-dimensional difference pair distribution function analysis (3D-$\Delta$PDF) to address this problem. 3D-$\Delta$PDFs visualise and quantify local deviations from the average structure and enable a straightforward interpretation of the single crystal diffuse scattering data in terms of a three-dimensional local order model. Comparison of the 3D-$\Delta$PDF from electron diffraction data with those obtained from neutron and x-ray experiments of yttria-stabilized zirconia (Zr$_{0.82}$Y$_{0.18}$O$_{1.91}$) demonstrates the reliability of the newly proposed approach.
			\end{abstract}

		\keywords{Diffuse scattering, 3D electron diffraction (3D ED), local order, microED, Yttria-stabilized zirconia (YSZ)}
\maketitle

	\section{Introduction}
	Functional materials design uses structure–property relationships which focus on structurally ordered systems, where disorder and defects are generally considered detrimental. However, it is known that certain types of correlated disorder can lead to phenomena that are inaccessible to ordered structures\cite{simonov_designing_2020,senn_emergence_2016,perversi_co-emergence_2019,weller_complete_2015}. Examples include the compositional complexity that drives domain structure in relaxor ferroelectrics (such as PbMg$_{1/3}$Nb$_{2/3}$O3)\cite{pasciak_polar_2012} and relaxor ferromagnets (such as LaNi$_{2/3}$Sb$_{1/3}$O$_{3}$)\cite{battle_3_2013} and Jahn-Teller distortions that give rise to specific electronic and magnetic properties as e.g., observed in LaMnO$_3$\cite{sartbaeva_quadrupolar_2007}.
	
	To successfully engineer correlated disorder in novel functional materials, the quantitative characterization of the local order is essential\cite{simonov_designing_2020,keen_crystallography_2015,welberry2016one}. In a diffraction experiment, long-range order manifests itself in Bragg reflections, the analysis of which is well established in conventional crystallography. Information about disorder is encoded in the much weaker diffuse scattering, the analysis of which is notoriously difficult and far from routine. Increasingly, powder pair distribution function analysis (powder PDF) from x-ray, neutron and/or electron scattering experiments have become the most powerful method for accessing local ordering motifs\cite{keen_crystallography_2015, proffen_structural_2003,keen_total_2020,mu_radial_2016} - especially in nanocrystalline samples that are not suitable for single crystal x-ray or neutron diffraction experiments. The major drawback of the powder PDF is the one-dimensional nature of the approach: the powder PDF essentially provides a histogram of interatomic distances, such that three-dimensional structural information is projected onto one dimension. This may lead to ambiguities if (1) interatomic-vectors in different directions have similar lengths or (2) if several (different) disordered phases are present as the inseparable ensemble average of all crystallites in the powdered samples is taken. 
	
	Three-dimensional single crystal diffuse scattering can be a remedy to resolve these ambiguities, as this approach (1) preserves the structural information in 3D space and (2) is performed on single crystals. There are many different approaches to the analysis of single crystal diffuse scattering\cite{welberry_diffuse_2022, neder2008diffuse,nield_diffuse_2001} of which the recently developed analysis of three-dimensional difference pair distribution function (3D-$\Delta$PDF\cite{weber_three-dimensional_2012,roth_solving_2019,simonov_yell:_2014}) provides the most intuitive and direct interpretation. As the name suggests, the 3D-$\Delta$PDF quantifies difference pair correlations. It is the Fourier transform of the diffuse scattering intensity without the Bragg scattering intensities. The correlations therefore correspond to local deviations away from the average structure, where positive correlations correspond to interatomic vectors with more scattering density in the real structure than suggested by the average structure, while negative correlations correspond to interatomic vectors with less scattering density than suggested by the average structure. An intuitive example would be the correlated thermal motion of neighbouring atoms: if neighbouring atoms vibrate in-phase, their interatomic vector is more confined than the average interatomic vector, which is smeared out by the average structure atomic displacement parameters (ADPs) assuming uncorrelated thermal motion. The resulting signature in the 3D-$\Delta$PDF is a sharp positive correlation surrounded by a region of negative correlations\cite{weber_three-dimensional_2012}.
	To date, 3D-$\Delta$PDF approaches have been successfully applied to single crystal x-ray and neutron diffraction measurements on various material classes\cite{simonov_experimental_2014,schmidt_direct_2023,krogstad_reciprocal_2020,guerin_elucidating_2022,meekel_truchet-tile_2023}. However, these methods require macroscopic single crystals, a requirement that is not always met for functional and applied materials, especially when properties of interest depend on the crystal size. In these cases, the analysis of larger samples from an adapted synthesis is not necessarily representative for the as-applied nano-sized material. Recent developments in the field of electron diffraction allow this limitation to be overcome: 3D electron diffraction (3D ED) experiments are routinely performed on submicron crystallites\cite{gemmi_3d_2019,gruene_establishing_2021}.
	
	Historically, single crystal electron diffuse scattering from submicron crystallites has been analysed using oriented zone axis patterns\cite{brink_electron_2002,withers_modulation_2015,withers_electron_1994,withers_oxygenfluorine_2003}, which only provide information in selected projections of the three-dimensional structure. Data acquisition using 3D ED methods can overcome this limitation\cite{poppe_quantitative_2022,krysiak_ab_2018,krysiak_new_2020} and we come to show how the resulting 3D data sets can be exploited in 3D-$\Delta$PDF analysis, which requires full reciprocal space coverage.
	
	In this article we demonstrate how the 3D-$\Delta$PDF technique can be applied to single crystal diffuse scattering data obtained from 3D ED data. The sample material chosen is yttria-stabilized zirconia (Zr$_{0.82}$Y$_{0.18}$O$_{1.91}$, YSZ) - a well-known and technologically important material for which large neutron size single crystals are readily available. By comparing the results of the 3D-$\Delta$PDF from x-ray and neutron experiments with our analysis of the electron diffraction data, we demonstrate the applicability and reliability of the proposed method.
	
\begin{figure}
	    	\centering
		\includegraphics[width=0.95\columnwidth]{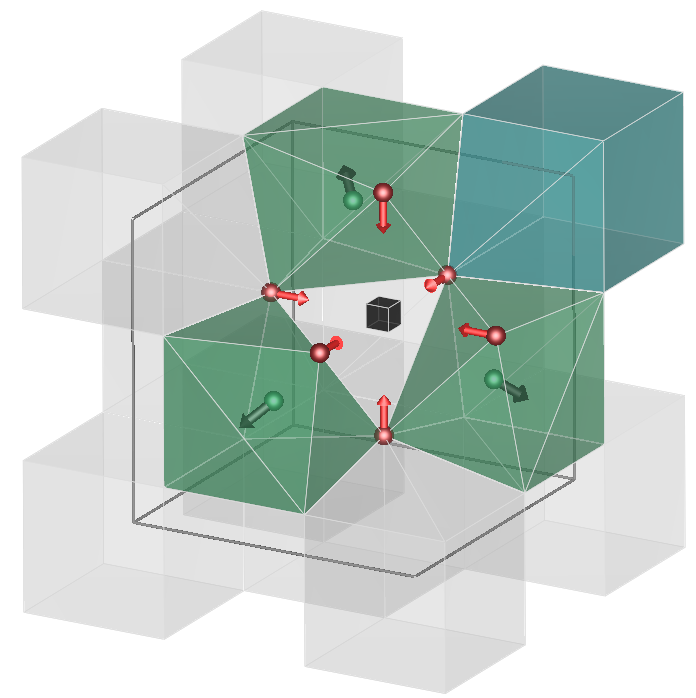}
		\caption{\textbf{Schematic defect model for cubic stabilized zirconia.} Grey polyhedra indicate Zr$^{4+}$ ions in 8-fold coordination. The solid black cube indicates a vacancy site. Zr$^{4+}$ ions in 7-fold coordination are displaced away from the vacancy along the $\langle 111\rangle$directions, indicated in green.  O$^{2-}$ ions directly next to the vacancy are displaced towards the vacancy along the $\langle 100\rangle$ directions, indicated in red. A possible Y$^{3+}$ site that is a next-nearest neighbour to the vacancy is indicated by the blue coordination polyhedron.
			}
		\label{fig1}
	\end{figure}

	YSZ shows pronounced diffuse scattering for which a consistent local order model has been established in the literature\cite{schmidt_direct_2023,welberry_3d_1993,frey_diffuse_2005,fevre_local_2005,khan_cation_1998}. Pure ZrO$_2$ adopts the cubic fluorite structure at elevated temperatures but is monoclinic at ambient conditions\cite{frey_diffuse_2005}. The cubic phase is stabilized at ambient conditions by the introduction of aliovalent oxides, such as Y$_2$O$_3$. The resulting compound adopts a disordered fluorite structure with the general formula Zr$_{(1-x)}$Y$_{x}$O$_{(2-x/2)}\square_{(x/2)}$, where $\square$ denotes a vacancy. The presence of oxygen vacancies is of great technological importance, as they are prerequisite for oxygen ion conduction\cite{tsampas_applications_2015}. From a structural point of view, oxygen vacancies lead to local distortions\cite{frey_diffuse_2005} as illustrated in Fig.~\ref{fig1}: oxygen ions neighbouring a vacancy shift along the $\langle 100 \rangle$ directions towards the vacancy, whereas metal ions neighbouring a vacancy shift along the $\langle 111\rangle$ directions away from the vacancy\cite{schmidt_direct_2023,frey_diffuse_2005,khan_cation_1998}. Furthermore, in YSZ there is a tendency to form 6-fold coordinated metal ions by vacancy pairs separated by $\langle \frac{1}{2} \frac{1}{2} \frac{1}{2} \rangle$ vectors\cite{schmidt_direct_2023,frey_diffuse_2005,khan_cation_1998}.
	
	These local correlations in YSZ lead to distinct signatures in the 3D-$\Delta$PDFs measured using x-ray and neutron diffraction. These signatures can be interpreted quantitatively in terms of a local order model\cite{schmidt_direct_2023}. Therefore, YSZ is an ideal reference material to establish and test the reliability of extracting and interpreting the electron 3D-$\Delta$PDF obtained from a submicron crystallite.
	
	\section{Results}
	
	\subsection{Reciprocal space}
	
\begin{figure*}[htb]
	\begin{minipage}[c]{0.61\textwidth}
		\includegraphics[width=\textwidth]{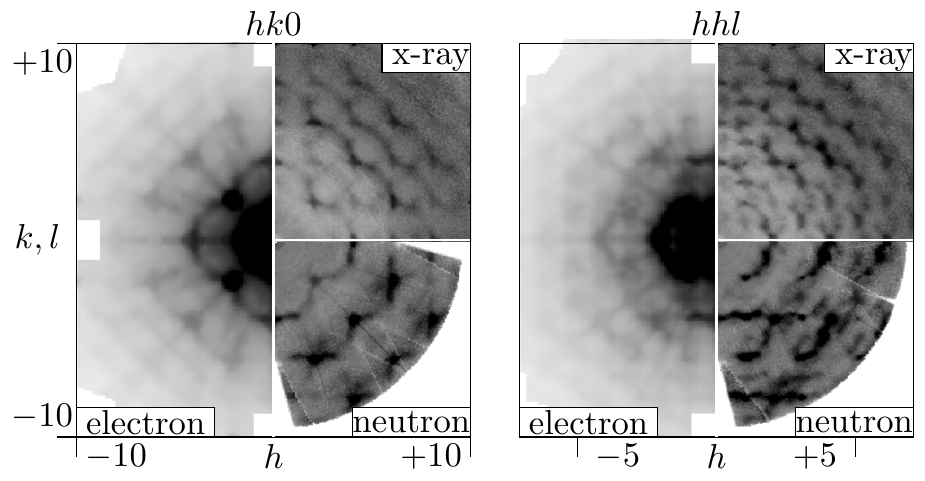}
		\end{minipage}
	\hfill
	\begin{minipage}[c]{0.27\textwidth}
		\caption{\textbf{Comparison of reciprocal space sections.} Diffuse scattering in reciprocal space sections after data treatment for x-ray, neutron and electron diffraction experiments. Left $hk0$-section, right $hhl$-section.
		}		\label{fig2}
	\end{minipage}

	\end{figure*}

	Electron diffraction patterns from an ion-milled single crystal of YSZ were measured with continuous-rotation 3D ED. Selected reciprocal space layers of single crystal diffuse scattering reconstructed from the experimentally obtained diffraction patterns are shown in Fig.~\ref{fig2} (for reconstruction and data processing routines see Methods section). The three probes chosen are sensitive to different structural aspects: in a diffraction experiment, neutrons probe exclusively the nuclei, x-rays probe almost exclusively the electron density and electrons probe the electrostatic potential, which depends on the distribution of both the nuclei and the electrons in the sample. Despite these differences, the main diffuse scattering features are similar in all three experiments. In the $hk0$-layer, for example, a continuously curved diffuse line connects the 400, 220 and 040 Bragg reflections, giving a flower-like pattern with higher order curved features at larger scattering angles. The $hhl$-layer shows streaks parallel to the $\langle110\rangle$ directions through the 004 Bragg reflection and a series of bracket-like features. One significant difference that can be observed is that the features in the electron diffraction data are broadened relative to the corresponding features in the x-ray and neutron data. This effect is seen also for the Bragg reflections, and so we attribute the broadening to our particular experimental setup.
	
\subsection{Real space}
	
			    \begin{figure*}
	\begin{minipage}[c]{0.7\textwidth}
	\includegraphics[width=\textwidth]{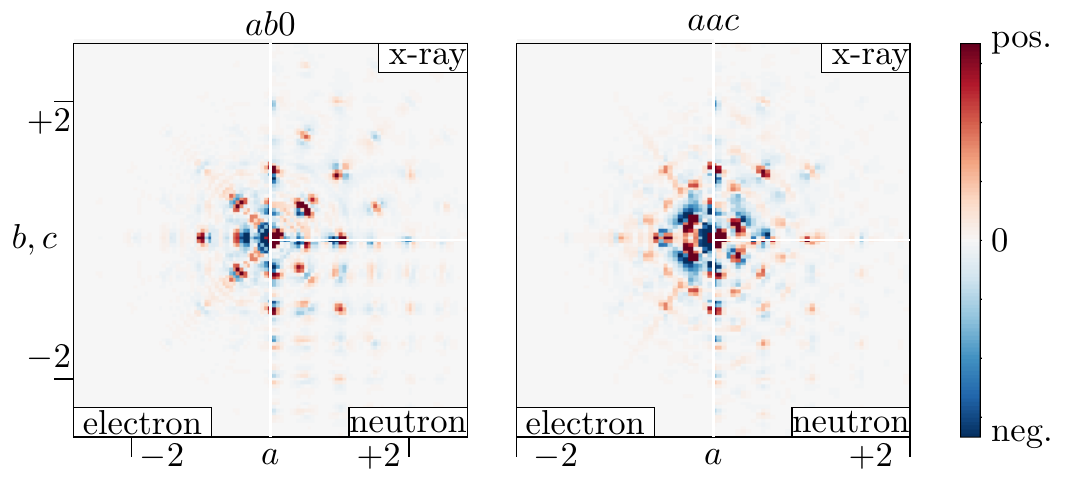}
\end{minipage}
\hfill
\begin{minipage}[c]{0.27\textwidth}
	\caption{\textbf{Comparison of 3D-$\Delta$PDF sections.} Sections generated from electron diffraction data compared to the corresponding sections from neutron and x-ray data. Left $ab0$-plane, right $aac$-plane. Positive correlations in red, negative correlations in blue.
	}
\label{fig3}
\end{minipage}

	\end{figure*}
	
	To develop a local order model from these data we use the 3D-$\Delta$PDF maps as shown in Fig \ref{fig3}. The first and most striking difference between the electron and other 3D-$\Delta$PDFs here is that the signatures in the electron 3D-$\Delta$PDF maps are much more localized in the centre of the PDF-space as compared to the x-ray and neutron maps. This is a direct consequence of the experimental broadening of the diffuse scattering in reciprocal space: the sharpness of diffuse scattering is inversely related to the correlation length of the corresponding local correlations in real space.
	
\begin{figure*}
		\begin{minipage}[c]{0.66\textwidth}
			\includegraphics[width=\textwidth]{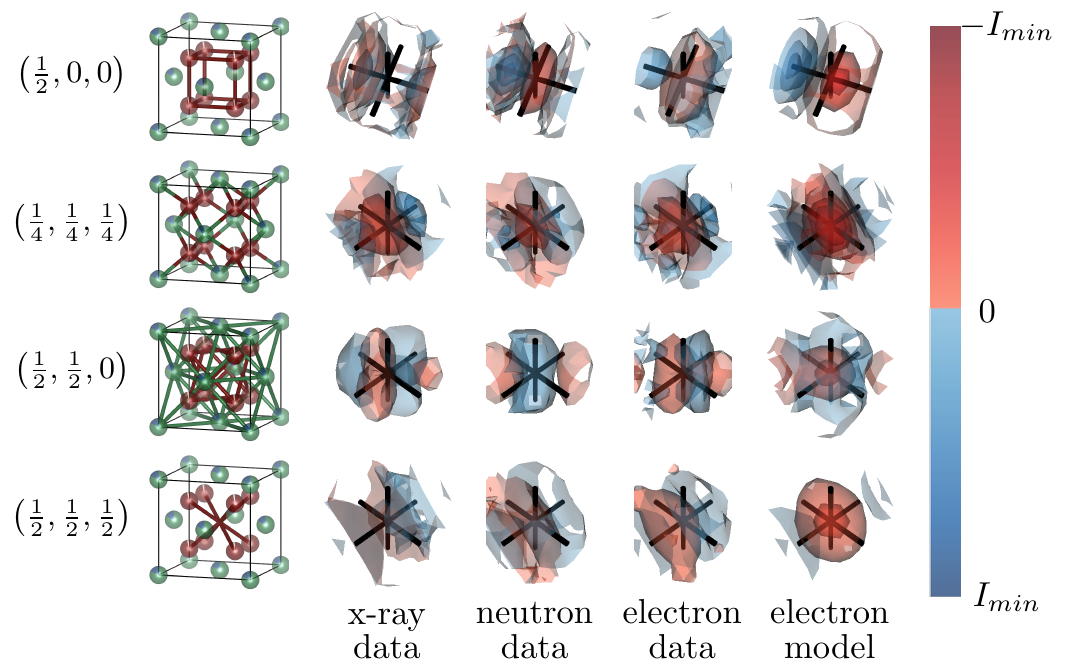}
		\end{minipage}
		\hfill
		\begin{minipage}[c]{0.3\textwidth}
			\caption{\textbf{Three-dimensional renderings of the 3D-$\Delta$PDF signatures at the shortest interatomic vectors generated from electron, neutron and x-ray diffraction experiments compared to the calculated signatures.} The leftmost column highlights the respective interatomic vectors as bonds in the average structure model. Zr$^{4+}$ in green, Y$^{3+}$ in blue, O$^{2-}$ in red. The rightmost column shows the signature generated from the calculated diffraction pattern a simplified model19 using the kinematic approximation. Black lines in the renderings are along the $\langle100\rangle$-directions and the crossing indicates the average interatomic vector. Rendering volume in the range $-0.15 \le \Delta a, \Delta b, \Delta c \le0.15$ around the average interatomic vector. Positive correlations in red, negative correlations in blue. Isosurfaces are shown relative to the minimum observed intensity in the respective 3D-$\Delta$PDF with the lowest Isosurface at 5 ~\% (1~\% for the $\left(\frac{1}{2},\frac{1}{2},\frac{1}{2}\right)$ interatomic vector).
			} 		\label{fig4}
		\end{minipage}

	\end{figure*}

	For a material with unknown local order, the goal is to identify the short-range deviations from the average structure and to model the very local interactions. The suppression of higher-order correlations does not hinder this goal. We compare the correlations at the shortest interatomic vectors in our 3D-$\Delta$PDF maps with those determined elsewhere using x-ray and neutron 3D-$\Delta$PDFs19,34 (see Figure 1). Three-dimensional renderings of the 3D-$\Delta$PDF regions that govern the local order model are shown in Fig. \ref{fig4}.
	
	The $\langle \frac{1}{2} 00\rangle$ interatomic vectors are the shortest interatomic vectors that correspond exclusively to difference vectors between oxygen positions. The signatures we observe in the electron, neutron and x-ray 3D-$\Delta$PDFs (see Fig.~\ref{fig4}) all show a minimum shifted by $\Delta_\mathrm{OO}^{-}$ along [100] towards the centre of PDF-space and a maximum shifted by $\Delta_\mathrm{OO}^{+}$ along [100] away from the centre of PDF-space. This is consistent with a local relaxation of oxygen ions towards neighbouring vacancy\cite{frey_diffuse_2005,schmidt_direct_2023}.
	
	The $\langle \frac{1}{4}\frac{1}{4}\frac{1}{4}\rangle$ interatomic vectors are the shortest interatomic vectors that correspond to difference vectors between oxygen and metal ion positions. In analogy to the $\langle \frac{1}{2} 00\rangle$ interatomic vectors, the signatures we observe from the three radiation types are consistent (see Fig. \ref{fig4}). All signatures show a minimum shifted by $\Delta_\mathrm{OM}^{-}$ along [111] away from the centre of PDF-space and a maximum shifted by $\Delta_\mathrm{OM}^{+}$  along [111]  towards the centre of PDF-space. This is consistent with a local contraction of the oxygen metal bond, indicating that metal ions adjacent to a vacancy shift away from that vacancy along the body diagonal\cite{schmidt_direct_2023}.
	
	The $\langle \frac{1}{2} \frac{1}{2} 0\rangle$ interatomic vectors are the shortest interatomic vectors that correspond to difference vectors between metal ion positions, but they also occur as difference vectors in the oxygen substructure. Thus, we observe a superposition of correlations resulting from metal-metal interactions and correlations resulting from oxygen-vacancy interactions \cite{schmidt_direct_2023}. This superposition is most pronounced in the neutron 3D-$\Delta$PDF, since the neutron scattering length of oxygen is comparable to that of the metals, whereas in x-ray and electron diffraction the metals dominate the scattering process and consequently the correlations in the 3D-$\Delta$PDFs (see Supporting Information for the quantitative comparison of the scattering factors). In the present case this allows us to assign the maximum observed at $\left(\frac{1}{2}+\Delta_\mathrm{MM}^{+},\frac{1}{2}1+\Delta_\mathrm{MM}^{+},0\right)$ in the x-ray and electron 3D-$\Delta$PDF to an elongated metal-metal vector resulting from two metal ions that are bridged by one oxygen ion and one vacancy, both shifting away from this vacancy \cite{schmidt_direct_2023}.
	
	\subsection{Quantitative comparison}
\begin{figure*}
		\begin{minipage}[c]{0.66\textwidth}
			\includegraphics[width=\textwidth]{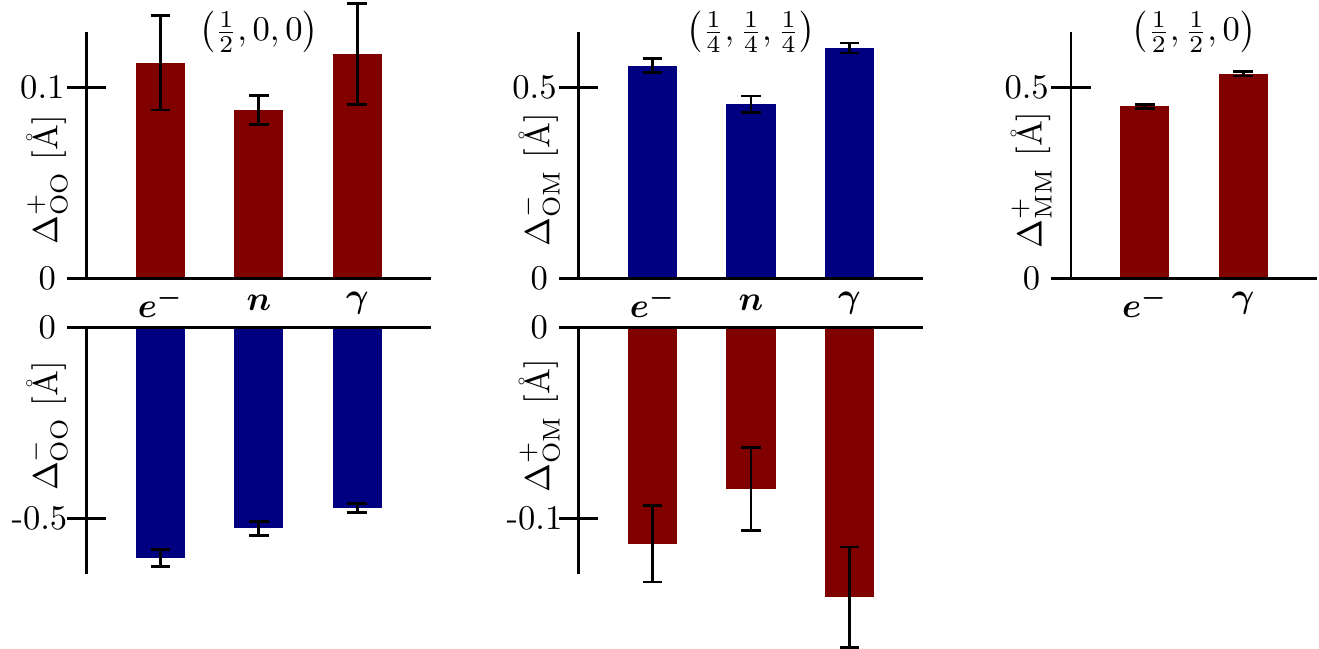}
		\end{minipage}
		\hfill
		\begin{minipage}[c]{0.3\textwidth}
			\caption{\textbf{Shifts of the observed maxima (red) and minima (blue) away from the average interatomic distance.} Shifts are estimated by fitting a three-dimensional Gaussian distribution to the data displayed in Fig.~\ref{fig4}. Positive shift magnitudes correspond to shifts away from the centre of real space, i.e., elongated interatomic distances. Error bars indicate the 3$\sigma$ level of the fit uncertainty. For details see Tables S1-3 in the supporting information..
			} 		\label{fig5}
		\end{minipage}

	\end{figure*}

To evaluate the quantitative reliability of the 3D-$\Delta$PDF, we quantify the shifts of the maxima and minima described qualitatively in the previous section ($\Delta_\mathrm{OO}^+$ , $\Delta_\mathrm{OO}^-$, $\Delta_\mathrm{OM}^+$ ,$\Delta_\mathrm{OM}^-$ , and $\Delta_\mathrm{MM}^+$). We approximate the intensity distributions at the shortest inter-atomic vectors with three-dimensional Gaussian distributions \cite{schmidt_direct_2023}. The refined shift magnitudes are visualized in Fig.~\ref{fig5}; a complete list of the parameters including the variances is given in Tables S1-S3 in the supporting information. What is immediately clear is that the refined quantities are similar for all three types of radiation used. This is a key result of our study and demonstrates the viability of electron 3D-$\Delta$PDF approaches.

Of course, we do not expect the refined shift magnitudes to be identical for all three radiation types used, as different probes are sensitive to different structural aspects. In particular, for the $\left(\frac{1}{4},\frac{1}{4},\frac{1}{4}\right)$ and $\left(\frac{1}{2},\frac{1}{2},0\right)$ vectors in PDF space, the shift magnitude corresponds to an average over the two metal ions. Because the different radiation types have different contrasts in the scattering length of Zr$^{4+}$ and Y$^{3+}$, differences in the quantified correlation vectors are therefore to be expected.

	\subsection{Increased relative sensitivity to lighter elements compared to x-ray diffraction data}
	
	The 3D-$\Delta$PDF analysis shows that the most pronounced local oxygen-oxygen interaction at the $\langle \frac{1}{2} 00\rangle$ interatomic vectors can be quantified using the x-ray 3D-$\Delta$PDF. The tendency of YSZ to form 6-fold coordinated metals with a pair of double vacancies separated by $\langle \frac{1}{2} \frac{1}{2}\frac{1}{2}\rangle$ vectors cannot be observed as a significant positive correlation in the x-ray 3D-$\Delta$PDF but can be clearly observed in the neutron 3D-$\Delta$PDF (compare Fig.~\ref{4}). The relative scattering strength of oxygen with respect to the metal ions is greater in electron diffraction than in x-ray diffraction (see supporting information for a graphical visualization). It is well known from conventional Bragg data analysis that this difference in scattering power enhances the detectability of the lightest elements in electron diffraction experiments compared to x-ray diffraction experiments\cite{palatinus_hydrogen_2017,klar_accurate_2023}. An excellent example of this relative sensitivity are the 3D-$\Delta$PDF signatures in the vicinity of the $\langle \frac{1}{2} \frac{1}{2}\frac{1}{2}\rangle$  interatomic vectors (compare Fig.~\ref{fig4}). In the neutron diffraction experiment we observe a weak but clear maximum at this interatomic vector. In the x-ray data, this signature is unresolved and indistinguishable from residual noise. But in the electron diffraction 3D-$\Delta$PDF, the same signature seen in the neutron data is now clearly identifiable. This observation highlights the increased sensitivity of the electron diffraction experiment to the lighter elements, which can be exploited not only in the analysis of Bragg data analysis but also in the analysis of diffuse scattering.
	
	\subsection{Comparison to a computational model}
	In an earlier study \cite{schmidt_direct_2023}, we constructed a local order model based on the x-ray and neutron diffraction data. This model was used as the basis for Monte Carlo simulations that generated an ensemble of $10\times10\times10$ supercells that captured the experimentally-determined local correlations in a simplistic atomistic model. With access to this model we were able to calculate the expected electron diffuse scattering (by applying the kinematic approximation), and hence the corresponding 3D-$\Delta$PDF \cite{neder2008diffuse}. Three-dimensional renderings of the calculated 3D-$\Delta$PDF around the shortest interatomic vectors are shown in the rightmost column of Fig.~\ref{fig4}; further two-dimensional sections that directly compare the calculated and experimentally obtained 3D-$\Delta$PDF are provided in the supporting information. The computational model reproduces the main features that we analyse using the 3D-$\Delta$PDF from the electron diffraction experiment as well as those from the x-ray and neutron experiments (compare \cite{schmidt_direct_2023}).
	
	\section{Discussion}
	
	\subsection{Reliability of the 3D-$\Delta$PDF from electron diffraction data}
	We have established that electron diffraction can yield sensible 3D-$\Delta$PDFs from submicron grains, overcoming the single-crystal limitations of x-ray and neutron techniques. The observed broadening of the diffuse scattering in the electron diffraction does not prevent a quantitative and reliable analysis. This is because the nature of local correlations is encoded in the position and shape of diffuse scattering, whereas the width of diffuse scattering features relates only to correlation length \cite{welberry_diffuse_2022,neder2008diffuse}. Thus, as we show here, it is possible to derive a local order model despite the experimentally observed broadening of the scattered intensities in the electron diffraction experiment. Determining such a local order model is usually the step considered most difficult in the analysis of diffuse scattering. The analysis of the extent of the correlations, which can be derived from the width of the observed features in reciprocal space, is straightforward and can in turn be solved by determining an instrumental resolution function in reciprocal space and hence the maximum observable correlation length \cite{goff_defect_1999}.
	
	Electrons interact much more strongly with matter than x-rays or neutrons. Therefore, the probability that a single electron passing through the sample will scatter more than once is significant. This effect is known as dynamical scattering and, among other things, breaks the inversion symmetry of reciprocal space \cite{palatinus_hydrogen_2017,klar_accurate_2023}. In our demonstration here we have treated the obtained diffraction data in a purely kinematic approach – hence ignoring dynamical scattering. Symmetry averaging for complete $m\bar{3}m$ Laue symmetry was applied. In the case of Bragg diffraction, the kinematic approximation has been routinely used to help solve and refine average structure models for more than a decade \cite{kolb_automated_2011,mugnaioli_ab_2009}; however, more accurate structure models are obtained by applying the dynamical theory of diffraction \cite{palatinus_hydrogen_2017, klar_accurate_2023, palatinus_structure_2015}. The effect of multiple scattering events on the 3D-$\Delta$PDF presents an obvious challenge to be addressed in future studies. Here, our quantitative analysis, the comparison to the x-ray and neutron experiments, and the agreement with a simple computational model calculated on a purely kinematic basis, collectively demonstrate that the kinematic approximation gives reasonable and useful results.
	
	\subsection{Challenges that need addressing}
	We show that it is possible to obtain a quantitative and reliable local order model from electron diffraction data using 3D-$\Delta$PDF analysis, but we recognise that many challenges – mostly experimental aspects - remain before this method can be used routinely.
	
	A major challenge that should be mentioned is the limited reciprocal space coverage available in electron diffraction experiments. The generation of the 3D-$\Delta$PDF from experimentally collected diffuse scattering data requires the application of a three-dimensional Fourier transform \cite{weber_three-dimensional_2012,simonov_yell:_2014}. Therefore, full reciprocal space coverage is required in the diffraction experiment. Modern single crystal x-ray and neutron diffractometers are designed so that this coverage is routinely achieved. This is not necessarily the case for electron diffraction experiments, which are mostly performed in conventional TEM setups, where only limited tilt angles are accessible (in our case the data were collected for one crystal and a goniometer tilt range of $\pm 50^\circ$). If the symmetry in reciprocal space is high, then the missing wedge in reciprocal space can be filled by symmetry averaging, as we have done here for YSZ (see Methods). However, this does not allow us to check whether the diffuse scattering really reproduces the average structure symmetry, and for lower symmetry structures this option may not be sufficient to achieve a full reciprocal space coverage. In this case, we suggest combining data from several crystals, which is the typical approach used to increase completeness in 3D ED \cite{wennmacher_3d-structured_2019}.
	
	Another challenge is the stability of the sample with respect to the electron-beam-induced radiation damage and vacuum exposure. Radiation damage may be an issue even in x-ray diffraction experiments \cite{teng_primary_2000,coates_negative_2021}, but samples in electron diffraction experiments are even more susceptible to radiation damage due to their small volume. If the material is beam-sensitive, only a small amount of data can be collected from a single crystal, resulting in incomplete data. The problem can be overcome somewhat by cooling the sample during data collection \cite{gemmi_3d_2019,mugnaioli_electron_2020}. Again, as in other cases of data incompleteness, combining data from several crystals may help. 
	
	A final point to consider is the method of detecting the diffuse scattering. As the diffuse scattering intensities are typically $10^3$ to $10^4$ times weaker than the Bragg scattering intensity, the dynamic range of the detector is of paramount importance \cite{welberry_diffuse_2022}. In neutron diffraction experiments the neutron flux is the limiting factor in obtaining reliable diffuse scattering data \cite{welberry_diffuse_2022}. In x-ray diffraction, large area single photon-counting hybrid pixel detectors are standard in modern synchrotron and laboratory diffractometers \cite{welberry_diffuse_2022} and allow fast and almost noise-free data collection. In electron diffraction, it is essential that the detector dynamic range is sufficient to collect reliable diffuse scattering data in the presence of strong Bragg reflections.
	
	\subsection{Prospective applications}
	Electron diffraction experiments are routinely performed on submicron crystallites, as these are often synthesis products, whereas large single crystals suitable for x-ray or neutron diffraction experiments often need to be grown by especially adapted synthesis methods \cite{spingler_thoughts_2012}. Such a synthesis could potentially alter the local order properties. The use of 3D-$\Delta$PDFs from electron diffraction experiments will allow the characterization and quantification of local order phenomena in novel functional disordered materials, without the need to adapt the synthesis for the diffraction experiment. Furthermore, the use of electrons as a probe in the diffraction experiments improves the simultaneous detectability of light and heavy elements as compared to x-ray diffraction \cite{palatinus_hydrogen_2017,klar_accurate_2023}. This is crucial for functional oxide materials, such as our reference material YSZ, where both metal and oxygen ions are disordered. In these cases, the use of electrons as a probe allows the detection of weaker correlations of lighter elements.
	
	Powder PDF is and will remain an important characterization method, but its reduced information content can lead to difficulties and ambiguities of interpretation that are avoided with 3D methods. A key advantage of powder PDF, however, is that it easily allows in-situ, in-operando and variable pressure experiments, which currently can only be implemented to a very limited extent in a 3D ED setup. We envisage that many studies will now benefit from combining both approaches. For example, by establishing a three-dimensional local order model from 3D ED in conditions accessible to electron diffraction can provide a starting point for the adaption of the model that then describes the PDF data measured at the conditions of interest. We consider this combination to be the optimal use of the proposed method and we see great potential in its application to solving complex disorder problems.
	
	\section{Methods}
	
	\subsection{Sample material and preparation}
The zirconia samples have a composition of Zr$_{0.82}$Y$_{0.18}$O$_{1.91}$, grown by the skull melting method, delivered by Djevahirdjan S. A., Monthey, Switzerland. The composition was confirmed by energy-dispersive X-ray spectroscopy (EDX) \cite{schmidt_direct_2023}. For neutron measurements the large, clear single crystals were cut with a diamond saw to cubes with an edge length of approximately 5~mm. For X-ray diffraction measurements the larger crystals were mechanically ground to a diameter of about 150~$\mu$m and polished. For electron diffraction measurements a small grain of the crystal was thinned into a wedge using a focused ion beam on an spi supplies\textregistered\ Omniprobe (3 post, Cu) grid to a final thickness of 40 to 60 nm.	
	
	\subsection{X-ray diffraction measurements}
	
	X-ray experiments were performed on a Rigaku Synergy S diffractometer equipped with an Eiger 1M detector using Mo radiation, (sin($\theta_\mathrm{max}$)/$\lambda$ =1.28 \AA$^{-1}$). To avoid possible fluorescence a threshold of 17.4 keV was used on the detector. Simple $\omega$-scans with 0.5$^{\circ}$ step widths and 120 s exposure time were taken. The crystal was kept at ambient conditions. 3D diffuse scattering data was reconstructed on a $501\times501\times501$ voxel reciprocal space grid ($-10\le h,k,l \le 10$) using the orientation matrix provided by CrysAlis Pro \cite{agilent_crysalis_2014} and custom Python scripts using Meerkat \cite{simonov_meerkat_2020}.
	
	\subsection{Neutron diffraction measurements}
	Neutron diffraction experiments were carried out at D19 instrument ($\lambda$ = 0.95 \AA, 0.1$^{\circ}$ steps, 80 s exposure per frame, sin($\theta_\mathrm{max}$)/$\lambda$ = 0.94 \AA$^{-1}$), ILL, Grenoble utilizing a 180° $\phi$-scan. 3D diffuse scattering data reconstruction utilized the orientation matrix as provided by Int3d \cite{katcho_int3d_2021} and a custom Python script. The data was reconstructed on a $501\times501\times501$ voxel reciprocal space grid ($-10\le h,k,l \le 10$).
	
	\subsection{Electron diffraction measurements}
	Electron diffraction experiments were carried out at a FEI Tecani G2 microscope using 200 keV electrons ($\lambda$ = 0.02508 \AA, sin($\theta_\mathrm{max}$)/$\lambda$ =1.24 \AA$^{-1}$) equipped with a hybrid-pixel detector ASI Cheetah (512$\times$512 pixels). The goniometer was tilted from $-50^{\circ}$ to $+50^{\circ}$. Diffraction patters were taken with 1.0 s exposure time and the crystal was rotated by 0.25$^{\circ}$ between the patterns. PETS2 \cite{palatinus_specifics_2019} was used to refine the orientation and centre of the frames and to export the 3D diffuse scattering reciprocal space map. The data was reconstructed on a $201\times201\times201$ voxel reciprocal space grid ($-10\le h,k,l \le 10$). Due to the lower detector resolution and the observed broadening of the diffracted features this smaller grid was chosen for the electron diffraction experiment as compared to the neutron and x-ray experiments.
	
	\subsection{Data treatment procedures}
	The reflection conditions for the $F$-centring were fulfilled in all cases and after careful inspection the data were symmetry-averaged in the $m\bar{3}m$ Laue symmetry. The general data processing procedure to obtain 3D-$\Delta$PDF from experimental data is described in \cite{koch_single_2021}. The experimentally obtained data were treated with the KAREN outlier rejection algorithm \cite{weng_k-space_2020} and additionally a custom punch-and-fill approach that interpolates the intensity in punched voxels was used to eliminate residual Bragg intensities. To avoid Fourier ripples the data were multiplied with a Gaussian falloff that smooths the edges of the measured reciprocal space sections (see \cite{weng_k-space_2020}). The fast Fourier transform algorithm as implemented in Meerkat \cite{simonov_meerkat_2020} was used to obtain 3D-$\Delta$PDF maps.

	\section*{Acknowledgements}
Reinhard B. Neder (Erlangen), Arianna Minelli (Oxford) and Marie-Hélène Lemée (Grenoble) are thankfully acknowledged for providing the sample material, assisting in sample preparation and conducting the neutron diffraction experiment. We thankfully acknowledge Mariana Klementova for the help in the preparation of the sample for electron diffraction. The work presented here was supported by the European Research Council (advanced grant 788144 to A.L.G.). 
CzechNanoLab project LM2023051 funded by MEYS CR is gratefully acknowledged for the financial support of the measurements and sample fabrication at LNSM Research Infrastructure.
	
	\bibliographystyle{naturemag}
	\bibliography{ePDF_new} 

\begin{thebibliography}{10}
\expandafter\ifx\csname url\endcsname\relax
  \def\url#1{\texttt{#1}}\fi
\expandafter\ifx\csname urlprefix\endcsname\relax\def\urlprefix{URL }\fi
\providecommand{\bibinfo}[2]{#2}
\providecommand{\eprint}[2][]{\url{#2}}

\bibitem{simonov_designing_2020}
\bibinfo{author}{Simonov, A.} \& \bibinfo{author}{Goodwin, A.~L.}
\newblock \bibinfo{title}{Designing disorder into crystalline materials}.
\newblock \emph{\bibinfo{journal}{Nature Reviews Chemistry}}
  \textbf{\bibinfo{volume}{4}}, \bibinfo{pages}{657--673}
  (\bibinfo{year}{2020}).

\bibitem{senn_emergence_2016}
\bibinfo{author}{Senn, M.}, \bibinfo{author}{Keen, D.}, \bibinfo{author}{Lucas,
  T.}, \bibinfo{author}{Hriljac, J.} \& \bibinfo{author}{Goodwin, A.}
\newblock \bibinfo{title}{Emergence of long-range order in {BaTiO$_3$} from
  local symmetry-breaking distortions}.
\newblock \emph{\bibinfo{journal}{Physical review letters}}
  \textbf{\bibinfo{volume}{116}}, \bibinfo{pages}{207602}
  (\bibinfo{year}{2016}).

\bibitem{perversi_co-emergence_2019}
\bibinfo{author}{Perversi, G.} \emph{et~al.}
\newblock \bibinfo{title}{Co-emergence of magnetic order and structural
  fluctuations in magnetite}.
\newblock \emph{\bibinfo{journal}{Nature communications}}
  \textbf{\bibinfo{volume}{10}}, \bibinfo{pages}{2857} (\bibinfo{year}{2019}).

\bibitem{weller_complete_2015}
\bibinfo{author}{Weller, M.~T.}, \bibinfo{author}{Weber, O.~J.},
  \bibinfo{author}{Henry, P.~F.}, \bibinfo{author}{Di~Pumpo, A.~M.} \&
  \bibinfo{author}{Hansen, T.~C.}
\newblock \bibinfo{title}{Complete structure and cation orientation in the
  perovskite photovoltaic methylammonium lead iodide between 100 and 352 {K}}.
\newblock \emph{\bibinfo{journal}{Chemical communications}}
  \textbf{\bibinfo{volume}{51}}, \bibinfo{pages}{4180--4183}
  (\bibinfo{year}{2015}).

\bibitem{pasciak_polar_2012}
\bibinfo{author}{Paściak, M.}, \bibinfo{author}{Welberry, T.~R.},
  \bibinfo{author}{Kulda, J.}, \bibinfo{author}{Kempa, M.} \&
  \bibinfo{author}{Hlinka, J.}
\newblock \bibinfo{title}{Polar nanoregions and diffuse scattering in the
  relaxor ferroelectric {PbMg$_{1/3}$Nb$_{2/3}$O$_3$}}.
\newblock \emph{\bibinfo{journal}{Phys. Rev. B}} \textbf{\bibinfo{volume}{85}},
  \bibinfo{pages}{224109} (\bibinfo{year}{2012}).

\bibitem{battle_3_2013}
\bibinfo{author}{Battle, P.~D.}, \bibinfo{author}{Evers, S.~I.},
  \bibinfo{author}{Hunter, E.~C.} \& \bibinfo{author}{Westwood, M.}
\newblock \bibinfo{title}{La$_{3}${Ni} $_{3}${SbO}$_{9}$: a {Relaxor}
  {Ferromagnet}}.
\newblock \emph{\bibinfo{journal}{Inorg. Chem.}} \textbf{\bibinfo{volume}{52}},
  \bibinfo{pages}{6648--6653} (\bibinfo{year}{2013}).

\bibitem{sartbaeva_quadrupolar_2007}
\bibinfo{author}{Sartbaeva, A.}, \bibinfo{author}{Wells, S.~A.},
  \bibinfo{author}{Thorpe, M.~F.}, \bibinfo{author}{Božin, E.~S.} \&
  \bibinfo{author}{Billinge, S. J.~L.}
\newblock \bibinfo{title}{Quadrupolar {Ordering} in {LaMnO$_3$} {Revealed} from
  {Scattering} {Data} and {Geometric} {Modeling}}.
\newblock \emph{\bibinfo{journal}{Phys. Rev. Lett.}}
  \textbf{\bibinfo{volume}{99}}, \bibinfo{pages}{155503}
  (\bibinfo{year}{2007}).

\bibitem{keen_crystallography_2015}
\bibinfo{author}{Keen, D.~A.} \& \bibinfo{author}{Goodwin, A.~L.}
\newblock \bibinfo{title}{The crystallography of correlated disorder}.
\newblock \emph{\bibinfo{journal}{Nature}} \textbf{\bibinfo{volume}{521}},
  \bibinfo{pages}{303--309} (\bibinfo{year}{2015}).

\bibitem{welberry2016one}
\bibinfo{author}{Welberry, T.~R.} \& \bibinfo{author}{Weber, T.}
\newblock \bibinfo{title}{One hundred years of diffuse scattering}.
\newblock \emph{\bibinfo{journal}{Crystallography Reviews}}
  \textbf{\bibinfo{volume}{22}}, \bibinfo{pages}{2--78} (\bibinfo{year}{2016}).

\bibitem{proffen_structural_2003}
\bibinfo{author}{Proffen, T.}, \bibinfo{author}{Billinge, S.},
  \bibinfo{author}{Egami, T.} \& \bibinfo{author}{Louca, D.}
\newblock \bibinfo{title}{Structural analysis of complex materials using the
  atomic pair distribution function—{A} practical guide}.
\newblock \emph{\bibinfo{journal}{Zeitschrift für Kristallographie-Crystalline
  Materials}} \textbf{\bibinfo{volume}{218}}, \bibinfo{pages}{132--143}
  (\bibinfo{year}{2003}).

\bibitem{keen_total_2020}
\bibinfo{author}{Keen, D.~A.}
\newblock \bibinfo{title}{Total scattering and the pair distribution function
  in crystallography}.
\newblock \emph{\bibinfo{journal}{Crystallography Reviews}}
  \textbf{\bibinfo{volume}{26}}, \bibinfo{pages}{141--199}
  (\bibinfo{year}{2020}).

\bibitem{mu_radial_2016}
\bibinfo{author}{Mu, X.}, \bibinfo{author}{Wang, D.}, \bibinfo{author}{Feng,
  T.} \& \bibinfo{author}{Kübel, C.}
\newblock \bibinfo{title}{Radial distribution function imaging by {STEM}
  diffraction: {Phase} mapping and analysis of heterogeneous nanostructured
  glasses}.
\newblock \emph{\bibinfo{journal}{Ultramicroscopy}}
  \textbf{\bibinfo{volume}{168}}, \bibinfo{pages}{1--6} (\bibinfo{year}{2016}).

\bibitem{welberry_diffuse_2022}
\bibinfo{author}{Welberry, T.~R.}
\newblock \emph{\bibinfo{title}{Diffuse {X}-ray scattering and models of
  disorder}}, vol.~\bibinfo{volume}{31} (\bibinfo{publisher}{Oxford University
  Press}, \bibinfo{year}{2022}).

\bibitem{neder2008diffuse}
\bibinfo{author}{Neder, R.~B.} \& \bibinfo{author}{Proffen, T.}
\newblock \emph{\bibinfo{title}{Diffuse Scattering and Defect Structure
  Simulations: A cook book using the program DISCUS}},
  vol.~\bibinfo{volume}{11} (\bibinfo{publisher}{OUP Oxford},
  \bibinfo{year}{2008}).

\bibitem{nield_diffuse_2001}
\bibinfo{author}{Nield, V.~M.} \& \bibinfo{author}{Keen, D.~A.}
\newblock \emph{\bibinfo{title}{Diffuse neutron scattering from crystalline
  materials}}, vol.~\bibinfo{volume}{14} (\bibinfo{publisher}{Oxford University
  Press}, \bibinfo{year}{2001}).

\bibitem{weber_three-dimensional_2012}
\bibinfo{author}{Weber, T.} \& \bibinfo{author}{Simonov, A.}
\newblock \bibinfo{title}{The three-dimensional pair distribution function
  analysis of disordered single crystals: {Basic} concepts}.
\newblock \emph{\bibinfo{journal}{Zeitschrift fur Kristallographie}}
  \textbf{\bibinfo{volume}{227}}, \bibinfo{pages}{238--247}
  (\bibinfo{year}{2012}).

\bibitem{roth_solving_2019}
\bibinfo{author}{Roth, N.} \& \bibinfo{author}{Iversen, B.~B.}
\newblock \bibinfo{title}{Solving the disordered structure of
  $\beta$-{Cu$_{2-x}$Se} using the three-dimensional difference pair
  distribution function}.
\newblock \emph{\bibinfo{journal}{Acta Crystallographica Section A: Foundations
  and Advances}} \textbf{\bibinfo{volume}{75}}, \bibinfo{pages}{465--473}
  (\bibinfo{year}{2019}).

\bibitem{simonov_yell:_2014}
\bibinfo{author}{Simonov, A.}, \bibinfo{author}{Weber, T.} \&
  \bibinfo{author}{Steurer, W.}
\newblock \bibinfo{title}{Yell: {A} computer program for diffuse scattering
  analysis via three-dimensional delta pair distribution function refinement}.
\newblock \emph{\bibinfo{journal}{Journal of Applied Crystallography}}
  \textbf{\bibinfo{volume}{47}}, \bibinfo{pages}{1146--1152}
  (\bibinfo{year}{2014}).

\bibitem{simonov_experimental_2014}
\bibinfo{author}{Simonov, A.}, \bibinfo{author}{Weber, T.} \&
  \bibinfo{author}{Steurer, W.}
\newblock \bibinfo{title}{Experimental uncertainties of three-dimensional pair
  distribution function investigations exemplified on the diffuse scattering
  from a tris-tert-butyl-1,3,5-benzene tricarboxamide single crystal}.
\newblock \emph{\bibinfo{journal}{Journal of Applied Crystallography}}
  \textbf{\bibinfo{volume}{47}}, \bibinfo{pages}{2011--2018}
  (\bibinfo{year}{2014}).

\bibitem{schmidt_direct_2023}
\bibinfo{author}{Schmidt, E.} \emph{et~al.}
\newblock \bibinfo{title}{Direct interpretation of the {X}-ray and neutron
  three-dimensional difference pair distribution functions
  ({3D}-{$\Delta$}{PDFs}) of yttria-stabilized zirconia}.
\newblock \emph{\bibinfo{journal}{Acta Crystallographica Section B: Structural
  Science, Crystal Engineering and Materials}} \textbf{\bibinfo{volume}{79}},
  \bibinfo{pages}{138--147} (\bibinfo{year}{2023}).

\bibitem{krogstad_reciprocal_2020}
\bibinfo{author}{Krogstad, M.~J.} \emph{et~al.}
\newblock \bibinfo{title}{Reciprocal space imaging of ionic correlations in
  intercalation compounds}.
\newblock \emph{\bibinfo{journal}{Nature materials}}
  \textbf{\bibinfo{volume}{19}}, \bibinfo{pages}{63--68}
  (\bibinfo{year}{2020}).

\bibitem{guerin_elucidating_2022}
\bibinfo{author}{Guérin, L.} \emph{et~al.}
\newblock \bibinfo{title}{Elucidating {2D} {Charge}-{Density}-{Wave} {Atomic}
  {Structure} in an {MX}–{Chain} by the {3D}-{$\Delta$}{Pair} {Distribution}
  {Function} {Method}}.
\newblock \emph{\bibinfo{journal}{ChemPhysChem}} \textbf{\bibinfo{volume}{23}},
  \bibinfo{pages}{e202100857} (\bibinfo{year}{2022}).

\bibitem{meekel_truchet-tile_2023}
\bibinfo{author}{Meekel, E.~G.} \emph{et~al.}
\newblock \bibinfo{title}{Truchet-tile structure of a topologically aperiodic
  metal–organic framework}.
\newblock \emph{\bibinfo{journal}{Science}} \textbf{\bibinfo{volume}{379}},
  \bibinfo{pages}{357--361} (\bibinfo{year}{2023}).

\bibitem{gemmi_3d_2019}
\bibinfo{author}{Gemmi, M.} \emph{et~al.}
\newblock \bibinfo{title}{{3D} electron diffraction: the nanocrystallography
  revolution}.
\newblock \emph{\bibinfo{journal}{ACS Central Science}}
  \textbf{\bibinfo{volume}{5}}, \bibinfo{pages}{1315--1329}
  (\bibinfo{year}{2019}).

\bibitem{gruene_establishing_2021}
\bibinfo{author}{Gruene, T.}, \bibinfo{author}{Holstein, J.~J.},
  \bibinfo{author}{Clever, G.~H.} \& \bibinfo{author}{Keppler, B.}
\newblock \bibinfo{title}{Establishing electron diffraction in chemical
  crystallography}.
\newblock \emph{\bibinfo{journal}{Nature Reviews Chemistry}}
  \textbf{\bibinfo{volume}{5}}, \bibinfo{pages}{660--668}
  (\bibinfo{year}{2021}).

\bibitem{brink_electron_2002}
\bibinfo{author}{Brink, F.~J.}, \bibinfo{author}{Withers, R.~L.} \&
  \bibinfo{author}{Norén, L.}
\newblock \bibinfo{title}{An electron diffraction and crystal chemical
  investigation of oxygen/fluorine ordering in niobium oxyfluoride,
  {NbO$_2$F}}.
\newblock \emph{\bibinfo{journal}{Journal of Solid State Chemistry}}
  \textbf{\bibinfo{volume}{166}}, \bibinfo{pages}{73--80}
  (\bibinfo{year}{2002}).

\bibitem{withers_modulation_2015}
\bibinfo{author}{Withers, R.~L.}
\newblock \bibinfo{title}{A modulation wave approach to the order hidden in
  disorder}.
\newblock \emph{\bibinfo{journal}{IUCrJ}} \textbf{\bibinfo{volume}{2}},
  \bibinfo{pages}{74--84} (\bibinfo{year}{2015}).

\bibitem{withers_electron_1994}
\bibinfo{author}{Withers, R.}, \bibinfo{author}{Thompson, J.},
  \bibinfo{author}{Xiao, Y.} \& \bibinfo{author}{Kirkpatrick, R.}
\newblock \bibinfo{title}{An electron diffraction study of the polymorphs of
  {SiO$_2$}-tridymite}.
\newblock \emph{\bibinfo{journal}{Physics and Chemistry of Minerals}}
  \textbf{\bibinfo{volume}{21}}, \bibinfo{pages}{421--433}
  (\bibinfo{year}{1994}).

\bibitem{withers_oxygenfluorine_2003}
\bibinfo{author}{Withers, R.~L.}, \bibinfo{author}{Welberry, T.~R.},
  \bibinfo{author}{Brink, F.~J.} \& \bibinfo{author}{Norén, L.}
\newblock \bibinfo{title}{Oxygen/fluorine ordering, structured diffuse
  scattering and the local crystal chemistry of {K$_3$MoO$_3$F$_3$}}.
\newblock \emph{\bibinfo{journal}{Journal of Solid State Chemistry}}
  \textbf{\bibinfo{volume}{170}}, \bibinfo{pages}{211--220}
  (\bibinfo{year}{2003}).

\bibitem{poppe_quantitative_2022}
\bibinfo{author}{Poppe, R.}, \bibinfo{author}{Vandemeulebroucke, D.},
  \bibinfo{author}{Neder, R.~B.} \& \bibinfo{author}{Hadermann, J.}
\newblock \bibinfo{title}{Quantitative analysis of diffuse electron scattering
  in the lithium-ion battery cathode material {Li}
  $_{{1.2}}${Ni}$_{{0.13}}${Mn}$_{{0.54}}$ {Co}$_{{0.13}}$ {O}$_{{2}}$}.
\newblock \emph{\bibinfo{journal}{IUCrJ}} \textbf{\bibinfo{volume}{9}},
  \bibinfo{pages}{695--704} (\bibinfo{year}{2022}).

\bibitem{krysiak_ab_2018}
\bibinfo{author}{Krysiak, Y.}, \bibinfo{author}{Barton, B.},
  \bibinfo{author}{Marler, B.}, \bibinfo{author}{Neder, R.~B.} \&
  \bibinfo{author}{Kolb, U.}
\newblock \bibinfo{title}{\textit{{Ab} initio} structure determination and
  quantitative disorder analysis on nanoparticles by electron diffraction
  tomography}.
\newblock \emph{\bibinfo{journal}{Acta Crystallogr A Found Adv}}
  \textbf{\bibinfo{volume}{74}}, \bibinfo{pages}{93--101}
  (\bibinfo{year}{2018}).

\bibitem{krysiak_new_2020}
\bibinfo{author}{Krysiak, Y.} \emph{et~al.}
\newblock \bibinfo{title}{New zeolite-like {RUB}-5 and its related hydrous
  layer silicate {RUB}-6 structurally characterized by electron microscopy}.
\newblock \emph{\bibinfo{journal}{IUCrJ}} \textbf{\bibinfo{volume}{7}},
  \bibinfo{pages}{522--534} (\bibinfo{year}{2020}).

\bibitem{welberry_3d_1993}
\bibinfo{author}{Welberry, T.}, \bibinfo{author}{Butler, B.},
  \bibinfo{author}{Thompson, J.} \& \bibinfo{author}{Withers, R.}
\newblock \bibinfo{title}{A {3D} model for the diffuse scattering in cubic
  stabilized zirconias}.
\newblock \emph{\bibinfo{journal}{Journal of Solid State Chemistry}}
  \textbf{\bibinfo{volume}{106}}, \bibinfo{pages}{461--475}
  (\bibinfo{year}{1993}).

\bibitem{frey_diffuse_2005}
\bibinfo{author}{Frey, F.}, \bibinfo{author}{Boysen, H.} \&
  \bibinfo{author}{Kaiser-Bischoff, I.}
\newblock \bibinfo{title}{Diffuse scattering and disorder in zirconia}.
\newblock \emph{\bibinfo{journal}{Zeitschrift fur Kristallographie}}
  \textbf{\bibinfo{volume}{220}}, \bibinfo{pages}{1017--1026}
  (\bibinfo{year}{2005}).

\bibitem{fevre_local_2005}
\bibinfo{author}{Fèvre, M.}, \bibinfo{author}{Finel, A.} \&
  \bibinfo{author}{Caudron, R.}
\newblock \bibinfo{title}{Local order and thermal conductivity in
  yttria-stabilized zirconia. {I}. {Microstructural} investigations using
  neutron diffuse scattering and atomic-scale simulations}.
\newblock \emph{\bibinfo{journal}{Phys. Rev. B}} \textbf{\bibinfo{volume}{72}},
  \bibinfo{pages}{104117} (\bibinfo{year}{2005}).

\bibitem{khan_cation_1998}
\bibinfo{author}{Khan, M.~S.}, \bibinfo{author}{Islam, M.~S.} \&
  \bibinfo{author}{Bates, D.~R.}
\newblock \bibinfo{title}{Cation doping and oxygen diffusion in zirconia: {A}
  combined atomistic simulation and molecular dynamics study}.
\newblock \emph{\bibinfo{journal}{Journal of Materials Chemistry}}
  \textbf{\bibinfo{volume}{8}}, \bibinfo{pages}{2299--2307}
  (\bibinfo{year}{1998}).

\bibitem{tsampas_applications_2015}
\bibinfo{author}{Tsampas, M.}, \bibinfo{author}{Sapountzi, F.} \&
  \bibinfo{author}{Vernoux, P.}
\newblock \bibinfo{title}{Applications of yttria stabilized zirconia ({YSZ}) in
  catalysis}.
\newblock \emph{\bibinfo{journal}{Catalysis Science \& Technology}}
  \textbf{\bibinfo{volume}{5}}, \bibinfo{pages}{4884--4900}
  (\bibinfo{year}{2015}).

\bibitem{palatinus_hydrogen_2017}
\bibinfo{author}{Palatinus, L.} \emph{et~al.}
\newblock \bibinfo{title}{Hydrogen positions in single nanocrystals revealed by
  electron diffraction}.
\newblock \emph{\bibinfo{journal}{Science}} \textbf{\bibinfo{volume}{355}},
  \bibinfo{pages}{166--169} (\bibinfo{year}{2017}).

\bibitem{klar_accurate_2023}
\bibinfo{author}{Klar, P.~B.} \emph{et~al.}
\newblock \bibinfo{title}{Accurate structure models and absolute configuration
  determination using dynamical effects in continuous-rotation {3D} electron
  diffraction data}.
\newblock \emph{\bibinfo{journal}{Nat. Chem.}}  (\bibinfo{year}{2023}).

\bibitem{goff_defect_1999}
\bibinfo{author}{Goff, J.~P.}, \bibinfo{author}{Hayes, W.},
  \bibinfo{author}{Hull, S.}, \bibinfo{author}{Hutchings, M.~T.} \&
  \bibinfo{author}{Clausen, K.~N.}
\newblock \bibinfo{title}{Defect structure of yttria-stabilized zirconia and
  its influence on the ionic conductivity at elevated temperatures}.
\newblock \emph{\bibinfo{journal}{Phys. Rev. B}} \textbf{\bibinfo{volume}{59}},
  \bibinfo{pages}{14202--14219} (\bibinfo{year}{1999}).

\bibitem{kolb_automated_2011}
\bibinfo{author}{Kolb, U.}, \bibinfo{author}{Mugnaioli, E.} \&
  \bibinfo{author}{Gorelik, T.}
\newblock \bibinfo{title}{Automated electron diffraction tomography–a new
  tool for nano crystal structure analysis}.
\newblock \emph{\bibinfo{journal}{Crystal Research and Technology}}
  \textbf{\bibinfo{volume}{46}}, \bibinfo{pages}{542--554}
  (\bibinfo{year}{2011}).

\bibitem{mugnaioli_ab_2009}
\bibinfo{author}{Mugnaioli, E.}, \bibinfo{author}{Gorelik, T.} \&
  \bibinfo{author}{Kolb, U.}
\newblock \bibinfo{title}{“{Ab} initio” structure solution from electron
  diffraction data obtained by a combination of automated diffraction
  tomography and precession technique}.
\newblock \emph{\bibinfo{journal}{Ultramicroscopy}}
  \textbf{\bibinfo{volume}{109}}, \bibinfo{pages}{758--765}
  (\bibinfo{year}{2009}).

\bibitem{palatinus_structure_2015}
\bibinfo{author}{Palatinus, L.}, \bibinfo{author}{Petříček, V.} \&
  \bibinfo{author}{Corrêa, C.~A.}
\newblock \bibinfo{title}{Structure refinement using precession electron
  diffraction tomography and dynamical diffraction: theory and implementation}.
\newblock \emph{\bibinfo{journal}{Acta Crystallographica Section A: Foundations
  and Advances}} \textbf{\bibinfo{volume}{71}}, \bibinfo{pages}{235--244}
  (\bibinfo{year}{2015}).

\bibitem{wennmacher_3d-structured_2019}
\bibinfo{author}{Wennmacher, J. T.~C.} \emph{et~al.}
\newblock \bibinfo{title}{{3D}-structured supports create complete data sets
  for electron crystallography}.
\newblock \emph{\bibinfo{journal}{Nat Commun}} \textbf{\bibinfo{volume}{10}},
  \bibinfo{pages}{3316} (\bibinfo{year}{2019}).

\bibitem{teng_primary_2000}
\bibinfo{author}{Teng, T.~y.} \& \bibinfo{author}{Moffat, K.}
\newblock \bibinfo{title}{Primary radiation damage of protein crystals by an
  intense synchrotron {X}-ray beam}.
\newblock \emph{\bibinfo{journal}{Journal of synchrotron radiation}}
  \textbf{\bibinfo{volume}{7}}, \bibinfo{pages}{313--317}
  (\bibinfo{year}{2000}).

\bibitem{coates_negative_2021}
\bibinfo{author}{Coates, C.~S.}, \bibinfo{author}{Murray, C.~A.},
  \bibinfo{author}{Boström, H.~L.}, \bibinfo{author}{Reynolds, E.~M.} \&
  \bibinfo{author}{Goodwin, A.~L.}
\newblock \bibinfo{title}{Negative {X}-ray expansion in cadmium cyanide}.
\newblock \emph{\bibinfo{journal}{Materials Horizons}}
  \textbf{\bibinfo{volume}{8}}, \bibinfo{pages}{1446--1453}
  (\bibinfo{year}{2021}).

\bibitem{mugnaioli_electron_2020}
\bibinfo{author}{Mugnaioli, E.} \emph{et~al.}
\newblock \bibinfo{title}{Electron {Diffraction} on {Flash}-{Frozen}
  {Cowlesite} {Reveals} the {Structure} of the {First} {Two}-{Dimensional}
  {Natural} {Zeolite}}.
\newblock \emph{\bibinfo{journal}{ACS Cent. Sci.}}
  \textbf{\bibinfo{volume}{6}}, \bibinfo{pages}{1578--1586}
  (\bibinfo{year}{2020}).

\bibitem{spingler_thoughts_2012}
\bibinfo{author}{Spingler, B.}, \bibinfo{author}{Schnidrig, S.},
  \bibinfo{author}{Todorova, T.} \& \bibinfo{author}{Wild, F.}
\newblock \bibinfo{title}{Some thoughts about the single crystal growth of
  small molecules}.
\newblock \emph{\bibinfo{journal}{CrystEngComm}} \textbf{\bibinfo{volume}{14}},
  \bibinfo{pages}{751--757} (\bibinfo{year}{2012}).

\bibitem{agilent_crysalis_2014}
\bibinfo{author}{{Agilent}}.
\newblock \bibinfo{title}{{CrysAlis} {PRO}} (\bibinfo{year}{2014}).
\newblock \bibinfo{note}{Published: Agilent Technologies Ltd, Yarnton,
  Oxfordshire, England}.

\bibitem{simonov_meerkat_2020}
\bibinfo{author}{Simonov, A.}
\newblock \bibinfo{title}{Meerkat} (\bibinfo{year}{2020}).
\newblock \bibinfo{note}{Published: Vesion 0.3.7}.

\bibitem{katcho_int3d_2021}
\bibinfo{author}{Katcho, N.~A.}, \bibinfo{author}{Cañadillas-Delgado, L.},
  \bibinfo{author}{Fabelo, O.}, \bibinfo{author}{Fernández-Díaz, M.~T.} \&
  \bibinfo{author}{Rodríguez-Carvajal, J.}
\newblock \bibinfo{title}{{Int3D}: {A} {Data} {Reduction} {Software} for
  {Single} {Crystal} {Neutron} {Diffraction}}.
\newblock \emph{\bibinfo{journal}{Crystals}} \textbf{\bibinfo{volume}{11}},
  \bibinfo{pages}{897} (\bibinfo{year}{2021}).

\bibitem{palatinus_specifics_2019}
\bibinfo{author}{Palatinus, L.} \emph{et~al.}
\newblock \bibinfo{title}{Specifics of the data processing of precession
  electron diffraction tomography data and their implementation in the program
  {PETS2}.}
\newblock \emph{\bibinfo{journal}{Acta Crystallographica Section B: Structural
  Science, Crystal Engineering and Materials}} \textbf{\bibinfo{volume}{75}},
  \bibinfo{pages}{512--522} (\bibinfo{year}{2019}).

\bibitem{koch_single_2021}
\bibinfo{author}{Koch, R.~J.} \emph{et~al.}
\newblock \bibinfo{title}{On single-crystal total scattering data reduction and
  correction protocols for analysis in direct space}.
\newblock \emph{\bibinfo{journal}{Acta Crystallographica Section A: Foundations
  and Advances}} \textbf{\bibinfo{volume}{77}}, \bibinfo{pages}{611--636}
  (\bibinfo{year}{2021}).

\bibitem{weng_k-space_2020}
\bibinfo{author}{Weng, J.} \emph{et~al.}
\newblock \bibinfo{title}{K-space algorithmic reconstruction ({KAREN}): a
  robust statistical methodology to separate {Bragg} and diffuse scattering}.
\newblock \emph{\bibinfo{journal}{Journal of Applied Crystallography}}
  \textbf{\bibinfo{volume}{53}}, \bibinfo{pages}{159--169}
  (\bibinfo{year}{2020}).

\end{thebibliography}
	
\end{document}